\documentstyle[epsfig,11pt]{article}
\begin{document}
%\draft
%\twocolumn[\hsize\textwidth\columnwidth\hsize\csname@twocolumnfalse\endcsnae
\title{Statistical linguistic study of DNA sequences}
\author{K. L. Ng$^{\dag}$\\Department of Information Management, Ling
Tung College \\Tai-chung, Nantun, Taiwan 408 \\ S. P.
Li$^{\ast}$\\Institute of Physics, Academia Sinica, Nankang\\ Taipei,
Taiwan 115\\}
%\address{$^1$Ling Tung College, Department of Information
%Management,Nantun, Taichung, Taiwan 408, R.O.C. }
%\address{$^2$Institute of Physics, Academia Sinica, Taipei, Taiwan
%115,R.O.C.}
%\date{August, 2002}
\maketitle
 
\begin{abstract}
A new family of compound Poisson distribution functions from statistical
linguistic is used to study the n-tuples and nucleotide composition features 
of DNA sequences.  The relative frequency distribution of the 6-tuples and 7-
tuples occurrence studies suggest that most of the DNA sequences follow the 
general shape of the compound Poisson distribution.  It is also noted 
that the $\chi$-square test indicated that some of the sequences 
follow this distribution with a reasonable level of goodness of fit.    
The compositional segmentation study fits quite well using this new family of 
distribution functions.  Furthermore, the absolute values of the relative 
frequency come out naturally from the linguistic model without ambiguity.  It 
is suggesting that DNA sequences are not random sequences and they could 
possibly have subsequence structures.\\
Keywords : DNA segmentation, Statistical linguistic, Compound Poisson
distribution, Jensen-Shannon divergence measure
\end{abstract}
 
PACS numbers: 87.10.+e, 87.14.Gg \\
%\vspace{2pc}]
 
\section{Introduction}

In recent years, the subject of bioinformatics has emerged as an active
research subject in biology and other fields such as physics.  Researchers 
are beginning to search for DNA words [1,2] and build up dictionaries for
genomes [3].  In doing so, people are more willing to regard information
stored in DNA sequences as a natural language from nature.  A lot of these
activities are indeed to employ statistical methods to study DNA sequences.  
In an early attempt, researchers \cite{man} used
the Zipf's law \cite{zip1} to study the statistical features that are
embedded in DNA sequences.  Other researchers \cite{bor} subsequently used 
different distributions to fit the rank of the word distributions in DNA
sequences and obtained better fit than the original Zipf plot.  
The Zipf's law was first proposed in
1932 when George Zipf made an empirical observation on some
statistical regularities of human writings which has become the
most prominent statement of statistical linguistic.  In his
original work, Zipf found that if the number of different words
in a given text were arranged in the order of their frequency of
usage, there would be an approximate mathematical relation between
the frequency of occurrence of each word and its rank in the list
of all the words used in the text that were ordered by decreasing
frequency.  He later pointed out that similar relations also hold
in other contexts \cite{zip2, zip3}.
 
Let us associate a particular word by an index $r$ equal to its
rank, and by $f(r)$ the normalized frequency of occurrence of that
word, i.e., the number of times it appears in the text divided by
the total number of words $N$.  Zipf's law states that there
is an approximate relation between $f(r)$ and
$r$
 
$$
\displaystyle
f(r) = \frac{A}{r^\alpha} \,\, ,
\eqno(1)
$$
where $\alpha (> 1)$ and $A$ are constants.  The above mathematical
relation was used [4] to study the statistical features of DNA
sequences where similar scaling behavior was found.  It was
however noted that for sequences composed of primarily coding
regions, the data were well fitted by a logarithmic function \cite{bor}.
And just like in the case of linguistic, the Zipf's law could
only account for a limited zone of the rank variable.
 
In the early days of quantitative linguistics, researchers, notably
Yule [10] had suggested that the mathematical relation (1) proposed
by Zipf was unsatisfactory.  He conjectured that the correct
distribution for word frequencies would be a compound Poisson
model.  However, there is no fitting of any mathematical distribution
law to the extensive data in his book.  Good [11] later proposed to
overcome Yule's objection by introducing a convergence factor into
the Riemann distribution which gives
$$
f(r) = A r^{-\alpha} \theta^{r} \,\, ,
\eqno(2)
$$
where $r \ge 1, 0 < \theta < 1, \alpha > 0$ and $A$ is the 
normalization constant in terms of $\theta$ and $\alpha$.  Neither
Good nor any other authors have fitted (2) to any real data except
for $\theta$ very close to one.  One should also note that the
primary reason to introduce $\theta$ was to achieve convergence and
secondary to improve the fit.  In [10], Good also proposed a
distribution function to fit two sets of data but were rather
poor for large values of $r$.  Researchers in the field have also
introduced other distribution functions to fit word frequencies.  
However, as pointed out by Herdan [11], the word frequency
distribution functions are characterized by the properties of
both combinability and divisibility without altering the essential
mathematical characteristics of the distribution function.  The only
distribution functions have these two properties are of the 
compound Poisson type.  With this as the starting point, 
Sichel [12] introduced a new
family of compound Poisson distribution functions to fit word
frequencies.  He \cite{sic2} also used this family of distribution functions
to fit sentence-length, which was first considered by Yule [10].
The fit in both cases were encouraging.
 
A natural question to ask is whether the quantitative studies
made in linguistics can be carried out in a similar fashion in
DNA sequences.  In particular, we would like to know if the
compound Poisson distribution functions introduced in the study
of quantitative linguistics are universal, in the sense that
they can be used to study human designed languages such as
the languages we use everyday
as well as the language used by nature---the information stored in
DNA sequences.  We will answer the above question by carrying out
quantitative studies in DNA sequences using these compound Poisson
distribution functions.  In section II, we give a brief review of
the family of compound Poisson distribution function used in this
paper, which was first introduced by Sichel.  In section III, we
use this family of Poisson distribution functions to study the
statistical features of the {\it word frequencies} in DNA sequences.
Section IV is a statistical study of the {\it sentence-length} in
DNA sequences.  Section V is the discussion and summary.
 
\section{The Mathematical Model}
 
In this section, we give a brief introduction of the mathematical
model that we use throughout this paper.  We follow mainly the
discussion by Sichel [8].
 
Let the total vocabulary that an author uses in his writing consist
of $V$ distinct words.  For each word in the vocabulary $V$, there is a
long-term probability of occurrence $\pi_1, \pi_2, ..., \pi_V$ where
$\pi_1 (\pi_V)$ refers to the word with lowest and highest probability
respectively.  Theoretically, we have
$$
0 < \pi_i < 1 \,\,\, ; \,\,\, \sum_i \pi_i = 1 \,\,\, .
\eqno(3)
$$
The probability of a specific word to appear $r$ times in a total word
count
of N tokens is given by,
$$
\phi(r|N) = \left( \begin{array}{c}N\\r\\ \end{array} \right) \int^1_0
\pi^r
(1-\pi)^{N-r} \psi(\pi) d\pi \,\, .
\eqno(4)
$$
Since all the $\pi_i$'s are small, we may replace the binomial by the
Poisson distribution function.  We can then write $\lambda = N\pi$
and Eq.(2) becomes
$$
\displaystyle
\phi(r|N) = \frac{1}{r!} \int^1_0 e^{-N\pi} (N\pi)^r \psi(\pi) d\pi
= \frac{1}{r!} \int^N_0 e^{-\lambda}\lambda^r f(\lambda) d\lambda \,\, .
\eqno(3)
$$
For mathematical convenience, one can substitute $N$ by infinity in
the second integral in Eq.(3) since the latter is negligibly small
between $N$ and infinity.
 
The choice of the mixing distribution $\psi(\pi)$, or $f(\lambda)$,
is crucial.  A set of distribution functions was first suggested in [7].
Later, Sichel expressed Good's mixing distribution function as
$$
f(\lambda) = \frac{1}{2} \frac{(2(1-\theta)^{1/2}/\alpha\theta)^\gamma}
{K_\gamma(\alpha(1-\theta)^{1/2})} \, \lambda^{\gamma-1}
\exp\{ -(\frac{1}{\theta}-1) \lambda - \frac{\alpha^2\theta}{4\lambda}\}
\,
,
\eqno(4)
$$
where $-\infty < \gamma < \infty \, , 0 < \theta < 1$ and $\alpha > 0$
are constants and $K_\gamma$ is the modified Bessel function of
the second kind of order $\gamma$.  In particular, if $\gamma
= -\frac{1}{2}$,
$f(\lambda)$ will become the so-called inverse Gaussian distribution
function, which has applications in many different areas [12].
Substituting Eq.(4) into Eq.(3) and
perform a Bessel function integration, one will obtain the corresponding
compound Poisson distribution function
$$
\phi(r) =
\frac{((1-\theta)^{1/2})^\gamma}{K_\gamma(\alpha(1-\theta)^{1/2})}
\frac{(\alpha\theta/2)^r}{r!} K_{r+\gamma}(\alpha) \, ,
\eqno(5)
$$
where $r \ge 0$.  This three parameter family of discrete distribution
functions is extremely powerful.  A number of known distribution
functions such as the Poisson, negative binomial, geometric, Yule,
Good and Riemann distribution functions are all special or limiting
forms of Eq.(5).  If the parameter $\gamma$ is made negative in Eq.(5),
an entirely new set of discrete distribution functions is generated.
 
In general, the parameter $\alpha$ characterizes the frequency
behavior for low values of $r$, whereas $\theta$ influences
the tail and $\gamma$ is important for the entire sweep of the
distribution function.  For the calculation of the individual
probabilities in Eq.(5), Sichel derived a very useful formula based
on the following Bessel recursion relation
$$
K_{\nu+1}(z) = \frac{2\nu}{z} K_\nu(z) + K_{\nu-1}(z) \, .
\eqno(6)
$$
Using this recursion relation, one can easily obtain the following
recursion relation for $\phi(r)$
$$
\phi(r) = \theta ( \frac{r+\gamma-1}{r} ) \phi(r-1) +
\frac{(\alpha\theta)^2}{4r(r-1)} \phi(r-2) \, .
\eqno(7)
$$
Thus, as one obtains the first two probabilities $\phi(0)$ and
$\phi(1)$ from Eq.(5), it is easy to calculate all other probabilities
from Eq.(7).  It is clear that the word frequency distribution
functions start at $r = 1$.
A zero truncation of the function in Eq.(5) yields
$$
\phi(r) = [((1-\theta)^{1/2})^{-\gamma} K_\gamma(\alpha(1-\theta)^{1/2})
- K_\gamma(\alpha)]^{-1} \frac{(\alpha\theta/2)^r}{r!}
K_{r+\gamma}(\alpha)
\, ,
\eqno(8)
$$
for $r \ge 0$.  This is the Sichel model for word frequencies in its
most general form.
 
\section{Word frequencies in DNA sequences}
 
In this section, we use the Sichel model to study the {\it word
frequencies} in DNA sequences.  In order to adapt the Sichel model
to the quantitative study of DNA sequences, the concept of word must
first be defined.  In the case of coding regions, the words are the
64 3-tuples which code for the amino acids, AAA, AAT, etc.  For
noncoding regions, the words are however unknown.  Therefore, it is
better to consider the word length $n$ as a free parameter and
perform analyses for $n$ from say, 3 to 8 as was done in [1].
The number of $n$-tuples will be $4^n$.  Thus, for $n = 6$, the number
of the 6-tuples will be 4096.  To obtain the word frequency for each
$n$-tuple, we will start from the first base pair of the DNA sequence
that is under study and progressively shift by 1 base with a
window of length $n$.  For a DNA sequence containing $L$ base pairs,
the total number of words will be $L - n + 1$.
 
To avoid any bias in DNA sequence selection, we performed analysis
of 13 sequences of eukaryotes mammals (GenBank accession codes are
HSMHCAPG,
HUMGHCSA, HUMHBB, HUMHDABCD, HUMHPRTB, HUMMMDBC, HUMNEUROF, HUMRETBLAS,
HUMTCRADCV, HUMVITDBP, MMBGCXD, MUSTCRA, RATCRYG), 3 sequences of
invertebrate (CEC07A9, CELTWIMUSC, DROABDB), the yeast chromosome III
sequence (SCCHRIII), 10 sequences of eukaryotic viruses (ASFV55KB, HE1CG,
HEHCMVCG, HEVZVXX, HS1ULR, HSECOMGEN, HSGEND, IH1CG, VACCG, VVCGAA), 7
sequences of
bacteria (BSGENR, ECO110K, ECOHU47, ECOUW82, ECOUW85, ECOUW87, ECOUW89),
and 2 sequences of phage (LAMCG, MLCGA).
 
In Sichel's model, $\phi(r)$ is the fraction of the total number
of words with a frequency $r$ of appearance in the article under study.
For example, $\phi(1)$ is the fraction of words among the total number
of words used that appear once in the article.
To implement our analysis using the Sichel model, we first record
the total number ($N$) of words ($n$-tuples) among the total number
of possible words that are used in the DNA sequence under study.
For each frequency of appearance, we record the total
number ($N(r)$) of words ($n$-tuples) that have such a frequency $r$ of
appearance in that DNA sequence.  We divide that number by $N$ and
call it $\phi(r)$ and then plot $\phi(r)$ against $r$.  $\chi^2$ test is
used to obtain the best fit of the data against $\phi(r)$ in Eq.(5).
 
Table I is the result of the $\chi^2$ test of the DNA sequences chosen
from different groups of species using the Sichel model.  For each of the
DNA sequences, we give the result for the 6-tuples and 7-tuples.
We give the best values for $\gamma, \alpha$ and $\theta$ in each
case, which is determined by minimizing the $\chi^2$ value, 
$$
\chi^2= \sum_{i} \frac{(N_i-n_i)^2}{n_i}
\eqno(9)
$$
where $N_i$'s are the observed values and $n_i$'s are the theoretical 
values.  It is interesting to note that in the case of the 7-tuples,
most of the DNA sequences can be fit using an inverse Gaussian
distribution (i.e. $\gamma= -\frac{1}{2}$).
Fig.1 is an illustration of the $\chi^2$ test of some of the DNA
sequences chosen from Table I.  We choose
one sequence from each group of DNA sequences that we studied.  In
most cases, $\phi(r)$ can be fit reasonably well.

\section{Sentence Length in DNA sequences}
 
To study the sentence length of DNA sequences, one needs to define
what a sentence is.  In linguistics, it is easy to identify what a
sentence is.  In the case of DNA sequences, what exactly a
sentence should be is unknown.  One can, for example, identify the
word clusters in [3] as DNA sentences.  In our study here, 
we proceed with the following strategy.  We divide a DNA sequence into 
segments in such a way as to maximize the compositional divergence between 
the resulting
DNA domains until a stopping criterion is reached.  We then
identify each segment as a sentence in the DNA sequence.  In our
analysis, we use the segmentation method which employs the Jensen-Shannon
divergence measure [15] to study the bacterial DNA sequence, Eco110K, as an 
example.  We should remind our reader that one can use any other segmentation 
methods to study the sentence length in DNA sequences.  The number of 
segments 
($N(r)$) of length ($r$) is then recorded.  We again divide $N(r)$ by the 
total
number of segments to obtain the relative frequency distribution of
segments for $r$ and plot it against $r$, which is shown in Fig.2 [13].  In 
Fig.2,
we present the results for $d_r = 0.55, 0.60$ and $0.65$ which
correspond to significance level about 76\%, 79\% and 81\% respectively.
All of the chosen $d_r$s follow an inverse Gaussian distribution.  Table II 
is the result of the $\chi^2$ test of the Eco110K DNA sequences using the 
Sichel model.

\section{Summary and Discussion}
 
In the above, we have introduced a family of compound Poisson
distribution functions to the statistical study of DNA sequences.
We have used the compound Poisson distribution functions to fit
both the $n$-tuple and segment distributions of the DNA sequences.
In both cases, we have obtained reasonable fits, both the shape
and the normalization.  More interestingly, the relative frequency
distribution of $n$-tuples and the compositional segmentation study
follow the inverse Gaussian distribution among different types of species
and the normalization of $\phi(r)$ of both the word frequencies and
compositional segmentation comes out naturally from the linguistic model
without ambiguity.  

In the early linguistics feature studies [4] of DNA sequences, people
have plotted the relative occurrence of DNA {\it words} against rank and
found power law behaviors.  It was later shown that [6] the Zipf 
distribution indeed fits very poorly in many cases and the Yule (eq.(2))
distribution can give a much better fit.  However, as we have mentioned 
earlier, the Yule distribution was suggested primarily for the reason
of convergence and most of the fits are made when $\theta$ approaches
one.  This is also true in the case of [6] though the fits are 
better than that of [4].  On the other hand, the distribution suggested
by Sichel has a much rigorous base.  It is based on the fact that only
compound Poisson distribution functions have the properties that
characterize the word frequency distribution function in linguistics
and has its rigorous derivation.  This model incorporates the 
characteristic features of linguistics and thus 
a fit using Sichel's model should be of more theoretical interest.
We have indeed shown that one can obtain reasonably good fits using
this model.   

As mentioned in the above section, one would want to
study the compositional segmentation by using segmentation methods
other than the one we used here.  One of such methods is suggested in
[17].  In [17], new different stopping criteria for segmenting DNA
sequences are introduced.  The size of the segments are plotted 
against the rank and the result indicates that the Zipf plot is 
different for different segmentation methods.  It would be interesting
to see how the Sichel model can fit for different segmentation
methods.  This would then establish the validity of using 
the Sichel model in the linguistic study of DNA sequences.  It would be 
inappropriate to conclude that our results imply that DNA
sequences have any resemblance to a natural language.  However, it does 
suggest that DNA sequences are not random sequences and they could possibly 
have subsequence structures [19].

\section{Acknowledgment}
The work of K.L. Ng is support by the ROC NSC grant NSC 91-2626-E275-001, and 
the Academia Sinica short term visiting program.

\begin{table}[h]
\begin{center}
{Table 1.  Summary of the frequency distributions for 6-tuples and 7-tuples}
\begin{tabular}{|l|l|l|l|l|l|l|l|l|} \hline
\multicolumn{2}{|c|}{}&
\multicolumn{4}{|c|}{6-tuples}&
\multicolumn{3}{|c|}{7-tuples} \\ \cline{1-9}
Species&GenBankcode&$P(\chi^2)$&$\gamma$&$\alpha$&$\theta$&$\gamma$&$\alpha$&$\theta$\\ \hline
Mammal&HSMHCAPG&&1.7&0.4&0.91& -0.5&3.2& 0.91\\ \hline
&HUMGHCSA&0.48&1.3&0.3& 0.93&  -0.5&3.5& 0.91\\ \hline
&HUMHBB  &    &1.1&0.4& 0.95&  -0.5&3.9& 0.91\\ \hline
&HUMHDABCD& & 1.3& 2.2& 0.91&  -0.5&2.5& 0.91\\ \hline
&HUMHPRTB & & 1.3& 0.4& 0.91&  -0.5&2.8& 0.91\\ \hline
&HUMMMDBC&  & 0.5& 3.9& 0.95&  -0.5&2.4& 0.93\\ \hline
&HUMNEUROF& & 0.9& 0.6& 0.97&   0.3&2.3& 0.91\\ \hline
&HUMRETBLAS&& 1.1& 0.1& 0.93&   0.5&2.2& 0.93\\ \hline
&HUMTCRADCV&& 0.9& 1.3& 0.97&   0.3&2.5& 0.91\\ \hline
&HUMVITDBP& & 1.1& 0.4& 0.93&  -0.5&3.1& 0.91\\ \hline
&MMBGCXD&   & 1.1& 0.4& 0.93&  -0.5&3.3& 0.91\\ \hline
&MUSTCRA&   & 1.3& 0.4& 0.95&  -0.5&4.4& 0.91\\ \hline
&RATCRYG&   & 1.3& 0.4& 0.91&  -0.5& 2.7& 0.91\\ \hline
Invertebrate&CEC07A9&& 0.5&4.5&0.93&-0.5&2.5&0.91\\ \hline
&CELTWIMUSC&0.62& 0.5& 3.9& 0.93&   -0.7&2.5& 0.91\\ \hline
&DROABDB&       & 1.5& 4.5& 0.91&   -0.5&3.4& 0.91\\ \hline
Yeast ChrIII&SCCHRIII&&2.3&2.1&0.97& 0.5& 4.5& 0.95\\ \hline
Virus&ASFV55KB& & 0.5& 4.5& 0.91&   -0.7&2.5&0.91\\ \hline
&HE1CG&     &0.5&4.5&0.97& -0.5&3.2& 0.97\\ \hline
&HEHCMVCG&  &2.9&2.5&0.95&  0.5&4.5& 0.93\\ \hline
&HEVZVXX&   &3.1&2.5&0.91& -0.5&4.5& 0.93\\ \hline
&HS1ULR&    &0.5&3.8&0.97& -0.5&3.1& 0.95\\ \hline
&HSECOMGEN& &2.5&4.5&0.93&  0.5&4.0& 0.91\\ \hline
&HSGEND&    &0.9&0.3&0.97&  0.7&0.1& 0.91\\ \hline
&IH1CG&     &2.5&0.1&0.93& -0.5&4.5& 0.95\\ \hline
&VACCG&     &1.3&1.0&0.97& -0.5&4.2& 0.97\\ \hline
&VVCGAA&    &1.3&0.4&0.97&  1.3&0.8&0.97\\ \hline
Bacteria&BSGENR&0.18&1.5&4.5&0.93& -0.5&4.1&0.91\\ \hline
&ECO110K&       0.24&2.7&0.2&0.91& -0.5&4.5&0.91\\ \hline
&ECOHU47&       0.26&1.5&4.5&0.91& -0.5&3.6&0.91\\ \hline
&ECOUW82&           &2.5&2.8&0.93&  0.5&3.2&0.91\\ \hline
&ECOUW85&           &2.1&2.5&0.91& -0.5&4.1&0.91\\ \hline
&ECOUW87&           &2.3&2.5&0.91&  0.5&4.4&0.91\\ \hline
&ECOUW89&           &3.1&0.4&0.93&  0.5&4.5&0.91\\ \hline
%Bacteriaphage&LAMCG&&0.5&4.5&0.91& 0.071&-0.7&2.5&0.91\\ \hline
%&MLCGA&    &         1.1&0.2&0.93& 0.026&-0.5&2.8&0.91\\ \hline
\end{tabular}
\end{center}
\end{table}

\begin{table}
\begin{center}
{Table 1.  Summary of the frequency distributions for 6-tuples and 7-tuples}
\begin{tabular}{|l|l|l|l|l|l|l|l|l|} \hline
\multicolumn{2}{|c|}{}&
\multicolumn{4}{|c|}{6-tuples}&
\multicolumn{3}{|c|}{7-tuples} \\ \cline{1-9}
Species&GenBankcode&$P(\chi^2)$&$\gamma$&$\alpha$&$\theta$&$\gamma$&$\alpha$&$\theta$\\ \hline
Bacteriaphage&LAMCG&&0.5&4.5&0.91& -0.7&2.5&0.91\\ \hline
&MLCGA&    &         1.1&0.2&0.93& -0.5&2.8&0.91\\ \hline
\end{tabular}
\end{center}
\end{table}

\begin{table}
\begin{center}
{Table 2.  The frequency distribution for segmentation}
\begin{tabular}{|l|l|l|l|} \hline
$d_r$&$\gamma$&$\alpha$&$\theta$\\ \hline
0.50&    -0.5& 4.2& 0.97\\ \hline
0.60&    -0.5& 4.1& 0.99\\ \hline
0.65&    -0.5& 4.5& 0.99\\ \hline
\end{tabular}
\end{center}
\end{table}

\begin{figure}[h]
\begin{center}
\leavevmode
\epsfxsize 5in
\epsfysize 7in
\epsfbox{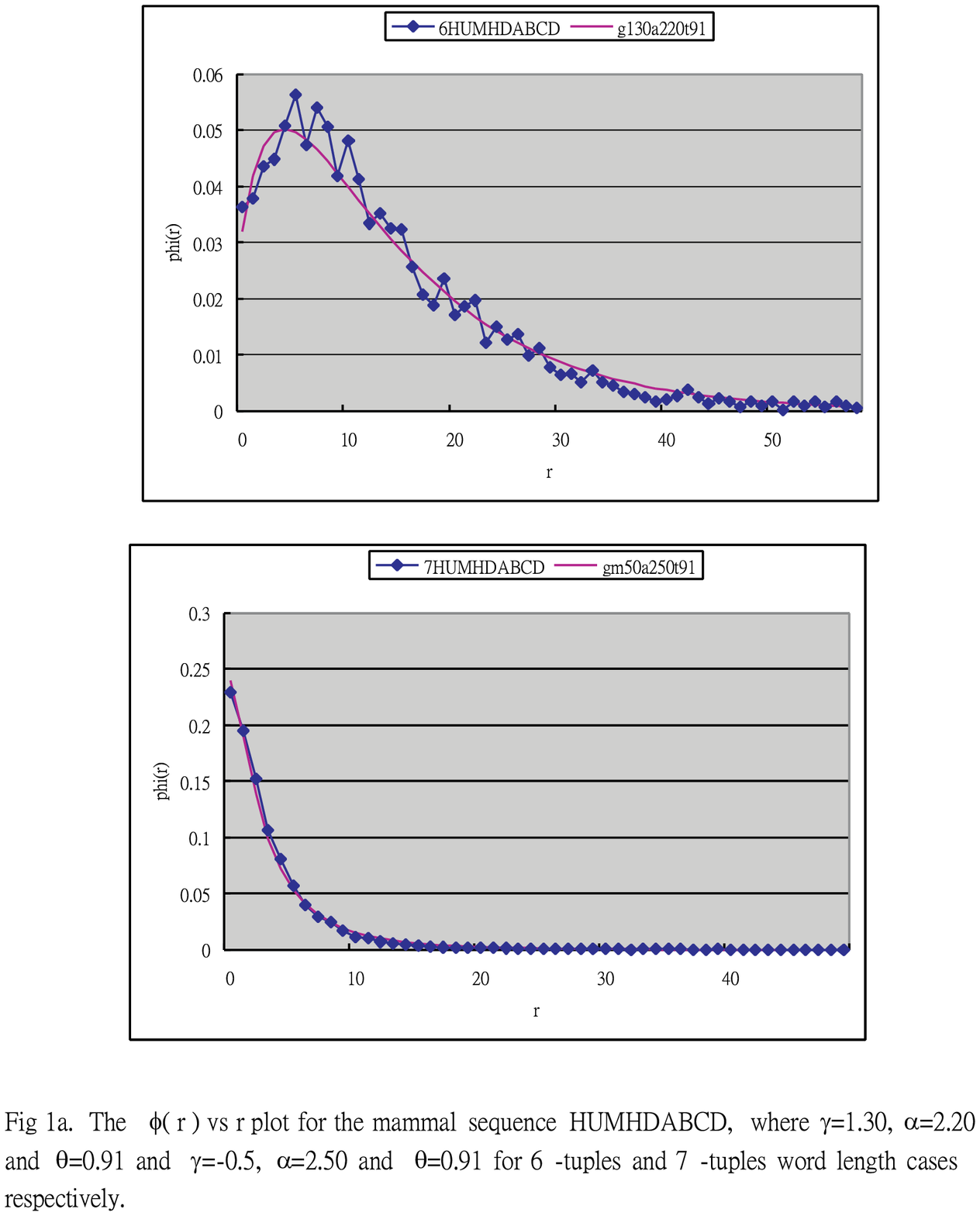}
\end{center}
\end{figure}

\begin{figure}[h]
\begin{center}
\leavevmode
\epsfxsize 5in
\epsfysize 7in
\epsfbox{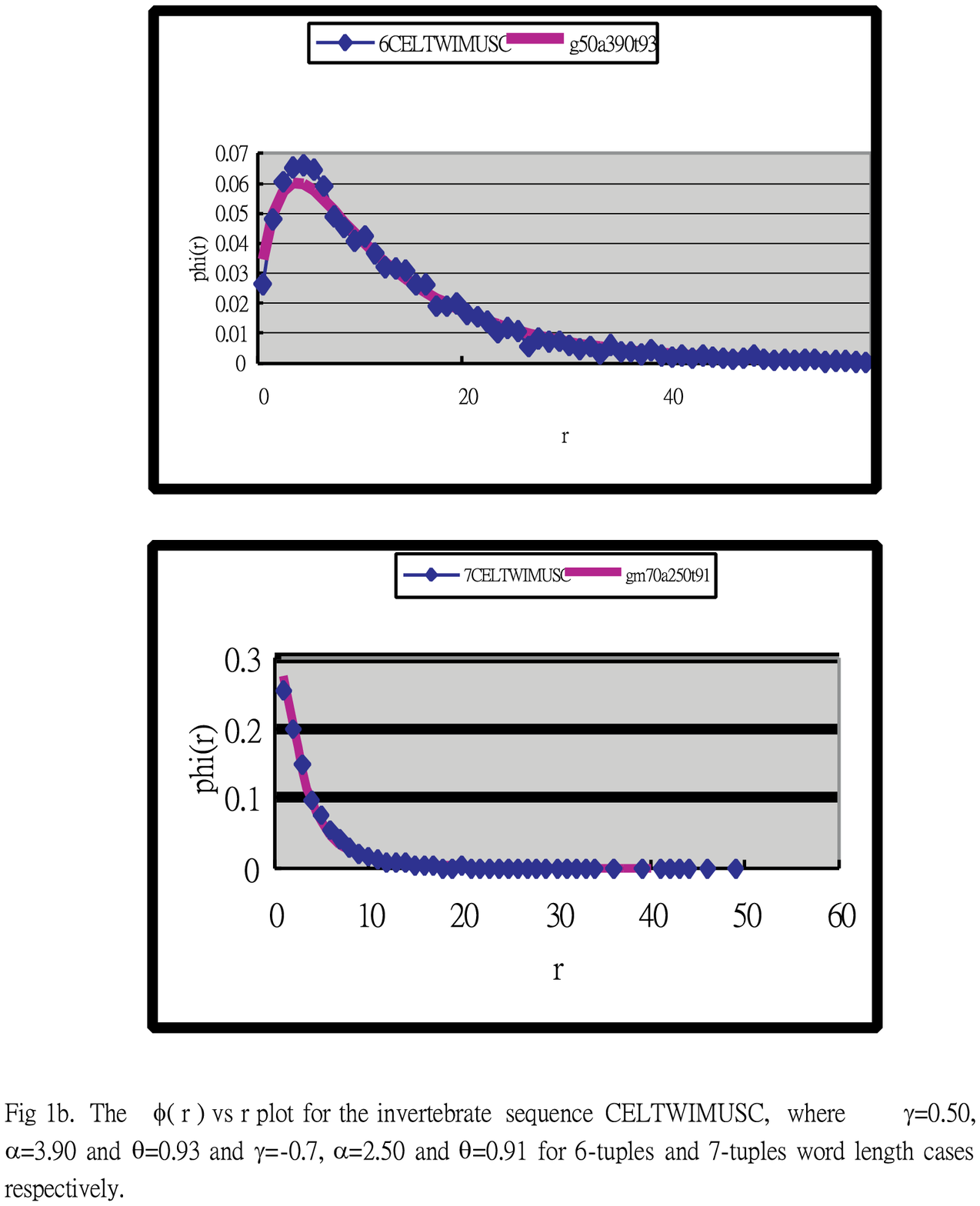}
\end{center}
\end{figure}

\begin{figure}[h]
\begin{center}
\leavevmode
\epsfxsize 5in
\epsfysize 7in
\epsfbox{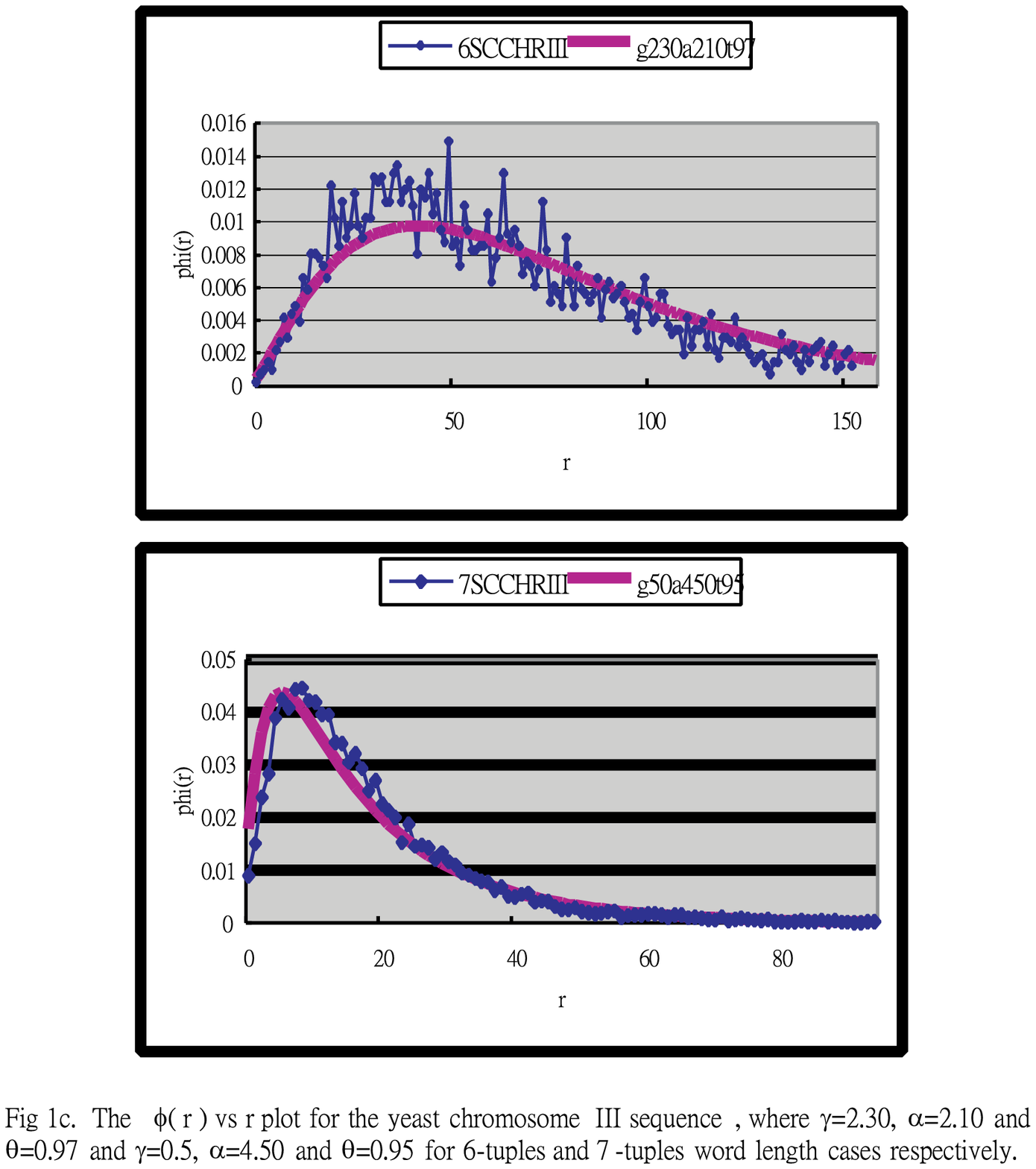}
\end{center}
\end{figure}

\begin{center}
\begin{figure}
\epsfxsize 5in
\epsfysize 7in
\epsfbox{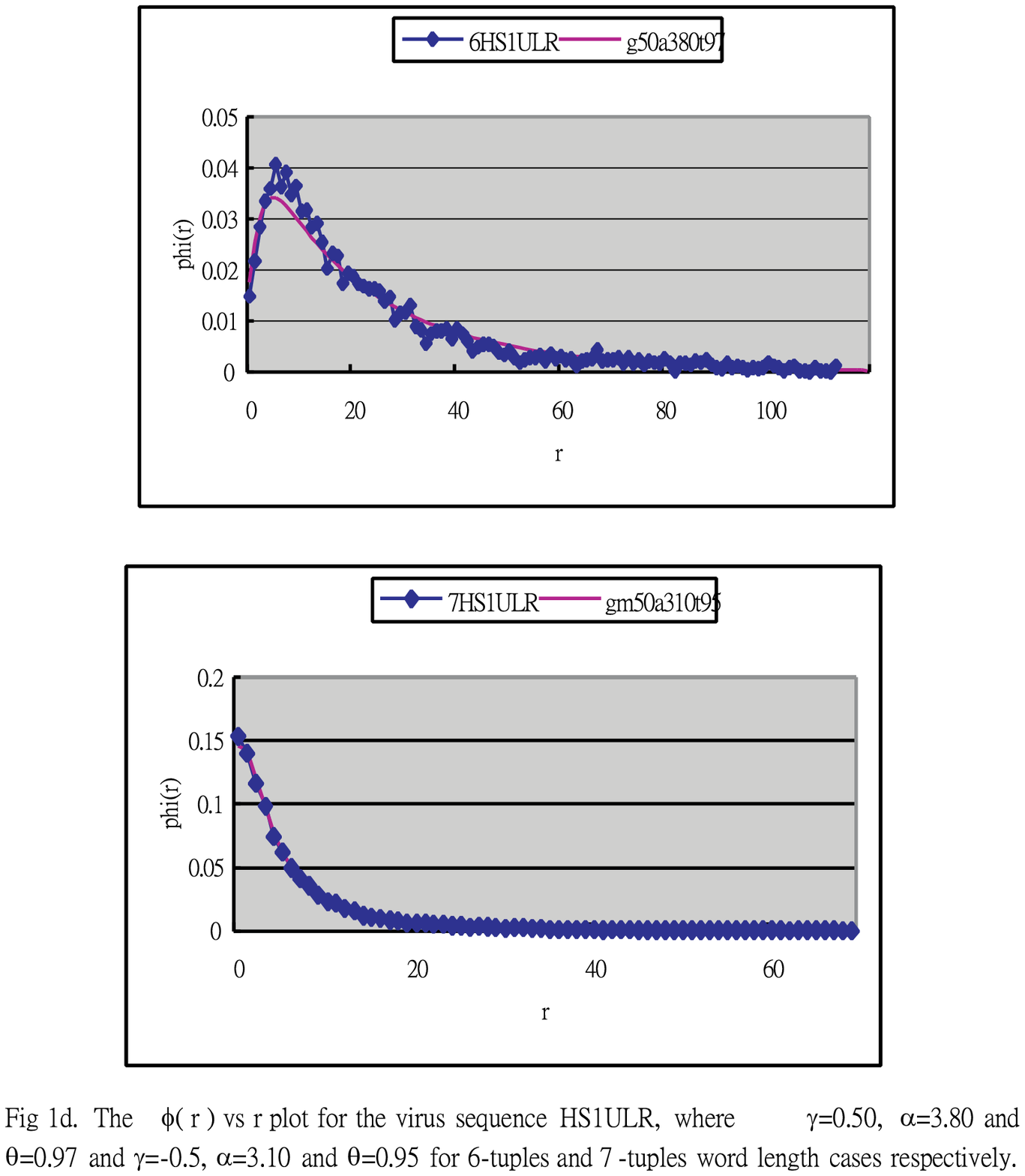}
\end{figure}
\end{center}

\begin{center}
\begin{figure}
\epsfxsize 5in
\epsfysize 7in
\epsfbox{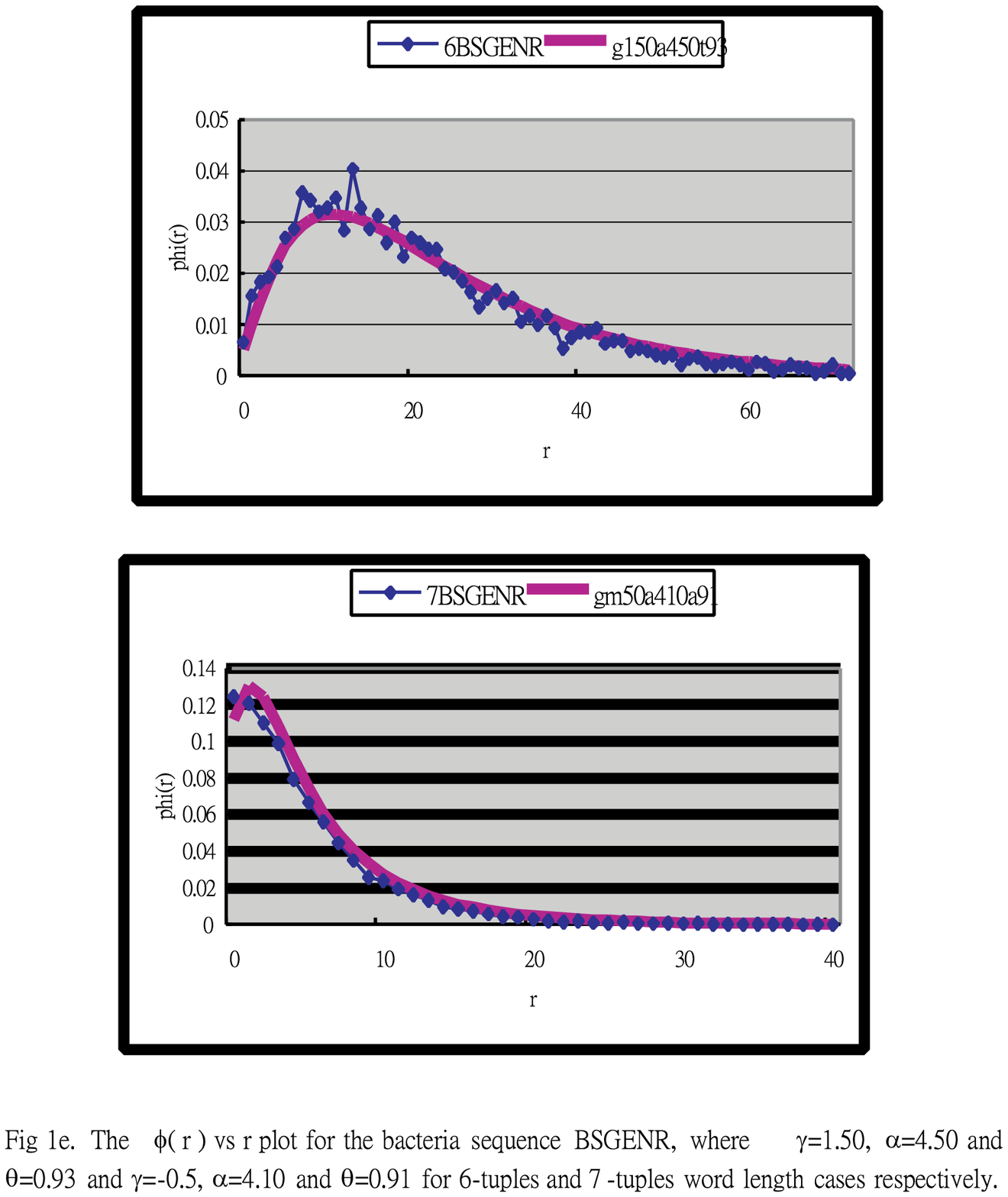}
\end{figure}
\end{center}

\begin{center}
\begin{figure}
\epsfxsize 5in
\epsfysize 7in
\epsfbox{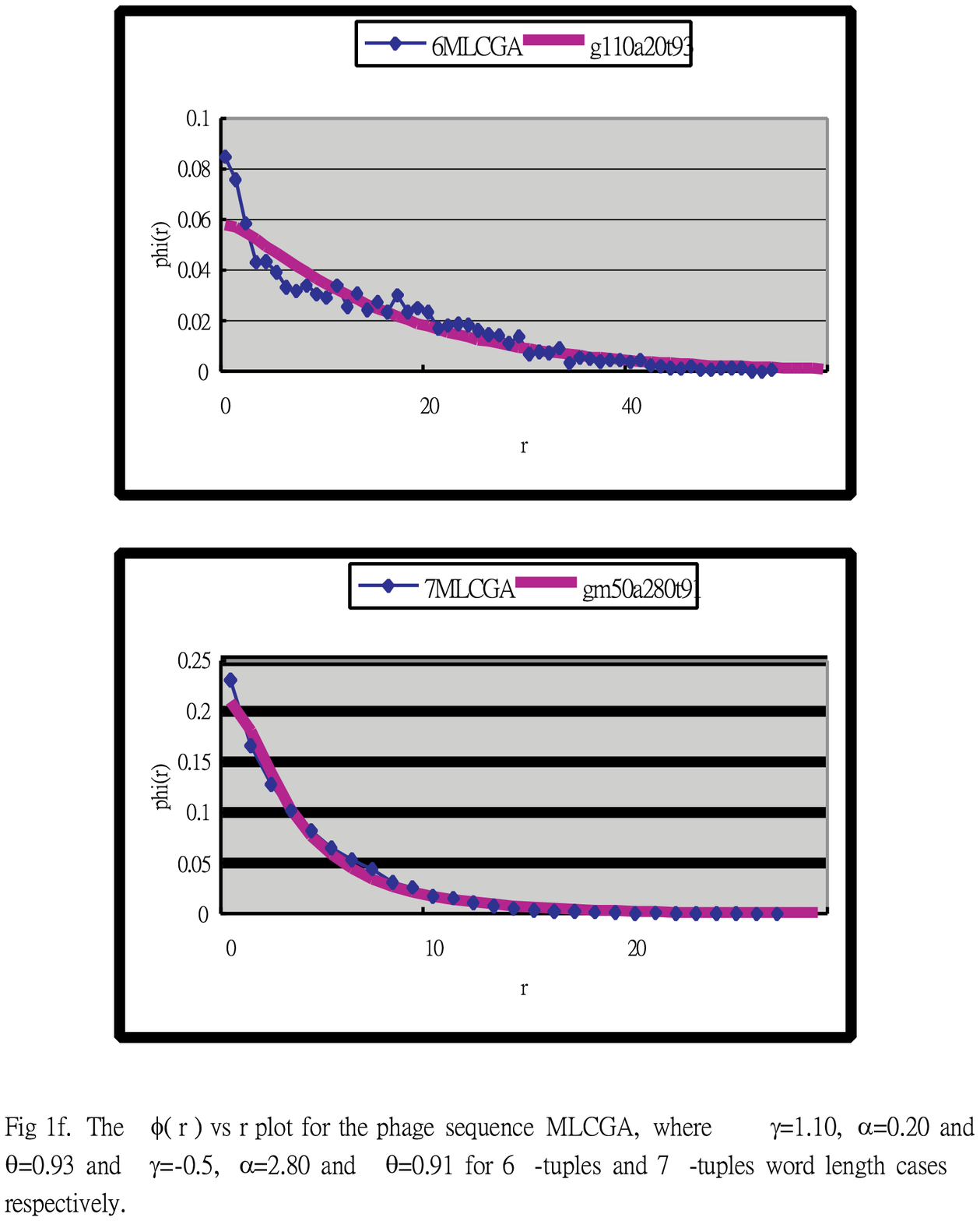}
\end{figure}
\end{center}

\begin{center}
\begin{figure}
\leavevmode
\epsfxsize 5in
\epsfysize 7in
\epsfbox{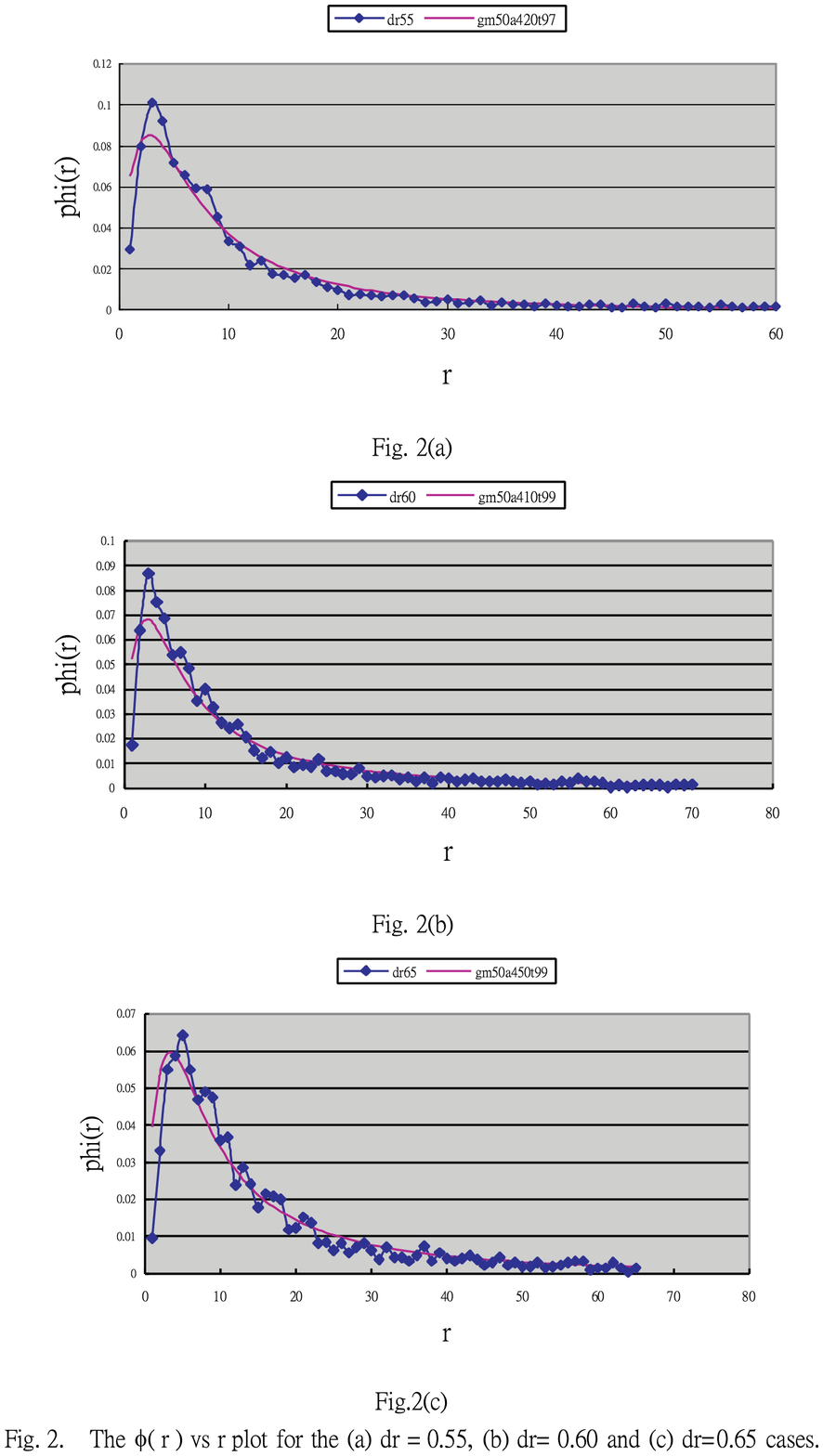}
\end{figure}
\end{center}

\end{document}